\documentclass[a4paper]{jpconf}
\usepackage{graphicx}

\newcommand{\affuni}[2]{Dipartimento di Fisica dell'Universit\`a #1, #2, Italy.}
\newcommand{\affinfn}[2]{INFN Sezione di #1, #2, Italy.}

\begin{document}

\title{\boldmath $U$ boson searches at KLOE}

\author{The KLOE-2 Collaboration}

\author{
F.~Archilli,$^{n,o}$
D.~Babusci,$^f$
D.~Badoni,$^{n,o}$
I.~Balwierz,$^e$
G.~Bencivenni,$^f$
C.~Bini,$^{l,m}$
C.~Bloise,$^f$
V.~Bocci,$^m$
F.~Bossi,$^f$
P.~Branchini,$^q$
A.~Budano,$^{p,q}$
S.~A.~Bulychjev,$^g$
P.~Campana,$^f$
G.~Capon,$^f$
F.~Ceradini,$^{p,q}$
P.~Ciambrone,$^f$
E.~Czerwi\'nski,$^f$
E.~Dan\'e,$^f$
E.~De~Lucia,$^f$
G.~De~Robertis,$^b$
A.~De~Santis,$^{l,m}$
G.~De~Zorzi,$^{l,m}$
A.~Di~Domenico,$^{l,m}$
C.~Di~Donato,$^{h,i}$
D.~Domenici,$^f$
O.~Erriquez,$^{a,b}$
G.~Fanizzi,$^{a,b}$
G.~Felici,$^f$
S.~Fiore,$^{l,m}$
P.~Franzini,$^{l,m}$
P.~Gauzzi,$^{l,m}$
S.~Giovannella,$^f$
F.~Gonnella,$^{n,o}$
E.~Graziani,$^q$
F.~Happacher,$^f$
B.~H\"oistad,$^s$
E.~Iarocci,$^{j,f}$
M.~Jacewicz,$^s$
T.~Johansson,$^s$
V.~Kulikov,$^g$
A.~Kupsc,$^s$
J.~Lee-Franzini,$^{f,r}$
F.~Loddo,$^b$
M.~Martemianov,$^g$
M.~Martini,$^{f,k}$
M.~Matsyuk,$^g$
R.~Messi,$^{n,o}$
S.~Miscetti,$^f$
G.~Morello,$^f$
D.~Moricciani,$^o$
P.~Moskal,$^e$
F.~Nguyen,$^{p,q}$
A.~Passeri,$^q$
V.~Patera,$^{j,f}$
I.~Prado~Longhi,$^{p,q}$
A.~Ranieri,$^b$
P.~Santangelo,$^f$
I.~Sarra,$^f$
M.~Schioppa,$^{c,d}$
B.~Sciascia,$^f$
A.~Sciubba,$^{j,f}$
M.~Silarski,$^e$
C.~Taccini,$^{p,q}$
L.~Tortora,$^q$
G.~Venanzoni,$^f$
R.~Versaci,$^{f,u}$
W.~Wi\'slicki,$^t$
M.~Wolke,$^s$
J.~Zdebik.$^e$
}

\begin{center}
$^a${\affuni{di Bari}{Bari}}\\
$^b${\affinfn{Bari}{Bari}}\\
$^c${\affuni{della Calabria}{Cosenza}}\\
$^d${INFN Gruppo collegato di Cosenza, Cosenza, Italy.}\\
$^e${Institute of Physics, Jagiellonian University, Cracow, Poland.}\\
$^f${Laboratori Nazionali di Frascati dell'INFN, Frascati, Italy.}\\
$^g${Institute for Theoretical and Experimental Physics (ITEP), Moscow, Russia.}\\
$^h${\affuni{''Federico II''}{Napoli}}\\
$^i${\affinfn{Napoli}{Napoli}}\\
$^j${Dipartimento di Scienze di Base ed Applicate per l'Ingegneria dell'Universit\`a ``Sapienza'', Roma, Italy.}\\
$^k${Dipartimento di Scienze e Tecnologie applicate, Universit\`a "Guglielmo Marconi", Roma, Italy.}\\
$^l${\affuni{''Sapienza''}{Roma}}\\
$^m${\affinfn{Roma}{Roma}}\\
$^n${\affuni{``Tor Vergata''}{Roma}}\\
$^o${\affinfn{Roma Tor Vergata}{Roma}}\\
$^p${\affuni{``Roma Tre''}{Roma}}\\
$^q${\affinfn{Roma Tre}{Roma}}\\
$^r${Physics Department, State University of New York at Stony Brook, USA.}\\
$^s${Department of Nuclear and Particle Physics, Uppsala Univeristy,Uppsala, Sweden.}\\
$^t${A. Soltan Institute for Nuclear Studies, Warsaw, Poland.}\\
$^u${Present Address: CERN, CH-1211 Geneva 23, Switzerland.}
\end{center}


\begin{abstract}
The existence of a secluded gauge sector could explain several puzzling 
astrophysical observations. This hypothesis can be tested at low energy
$e^+e^-$ colliders such as DA$\Phi$NE.
Preliminary results obtained with KLOE data and perpectives for the KLOE-2 
run, where a larger data sample is expected, are discussed.

\end{abstract}

\section{Introduction}

In recent years, several astrophysical observations have failed to find easy 
interpretations in terms of standard astrophysical and/or particle physics 
sources. A non exhaustive list of these observations includes the 511 keV 
gamma-ray signal from the galactic center observed by the INTEGRAL satellite
\cite{Jean:2003ci}, the excess in the cosmic ray positrons reported by 
PAMELA~\cite{Adriani:2008zr}, the total electron and positron flux measured 
by ATIC~\cite{Chang:2008zzr}, Fermi~\cite{Abdo:2009zk}, and HESS 
\cite{Collaboration:2008aaa,Aharonian:2009ah}, and the annual modulation of 
the DAMA/LIBRA signal~\cite{Bernabei:2005hj,Bernabei:2008yi}.     

An intriguing feature of these observations is that they suggest the 
existence of a WIMP dark matter particle belonging to a secluded gauge 
sector under which the Standard Model (SM) particles are uncharged
\cite{Pospelov:2007mp,ArkaniHamed:2008qn,Alves:2009nf,Pospelov:2008jd,%
Hisano:2003ec,Cirelli:2008pk,MarchRussell:2008yu,Cholis:2008wq,%
Cholis:2008qq,ArkaniHamed:2008qp}.
An abelian gauge field, the $U$ boson with mass near the GeV scale, 
couples the secluded sector to the SM through its kinetic mixing with 
the SM hypercharge gauge field.
The kinetic mixing parameter, $\epsilon$, can naturally be of the order 
10$^{-4}$--10$^{-2}$. 
In a very minimal scenario, in addition to the $U$, it is natural to have 
a secluded Higgs boson, the $h'$, which spontaneously breaks the secluded 
gauge symmetry. 
A consequence of the above hypotheses is that observable effects can be 
induced in $\mathcal{O}(\mbox{GeV}$)--energy $e^+e^-$ colliders
\cite{Batell:2009yf,Essig:2009nc,Reece:2009un,Bossi:2009uw,%
Borodatchenkova:2005ct,Yin:2009mc} and fixed target experiments 
\cite{Bjorken:2009mm,Batell:2009di,Essig:2010xa,Freytsis:2009bh}. 

\section{Searches for dark forces at KLOE}

The KLOE experiment operates at DA${\Phi}$NE, the $e^+e^-$ Frascati 
$\phi$-factory. From 2000 to 2006, KLOE collected 2.5 fb$^{-1}$ of
collisions at the $\phi$ meson peak and about 240 pb$^{-1}$ below the 
$\phi$ resonance ($\sqrt{s}=1$ GeV).
The $\phi$ meson predominantly decays into charged and neutral kaons,  
thus allowing KLOE to make precision studies in the fields of flavor 
physics, low energy QCD and test of discrete symmetries \cite{Cimento}. 

A new beam crossing scheme allowing a reduced beam size and increased 
luminosity is operating at DA$\Phi$NE \cite{CrabWaist}. 
The KLOE-2 detector was successfully installed in this new interaction 
region and has been upgraded with small angle tagging devices to detect 
both high and low energy electrons or positrons in $e^+e^-\to e^+e^-X$ 
events. About 5 fb$^{-1}$ are expected in the first year of running.
An inner tracker and small angle calorimeters are scheduled to be installed 
in a subsequent step, providing larger acceptance both for charged 
particles and photons.
A detailed description of the KLOE-2 physics program can be found in 
Ref.~\cite{KLOE2}.

The KLOE detector consists of a large cylindrical drift chamber, DC, 
surrounded by a lead-scintillating fiber electromagnetic calorimeter, 
EMC. A superconducting coil around the EMC provides a 0.52 T magnetic 
field. The all-stereo drift chamber~\cite{DCH}, 4~m in diameter and 
3.3~m long, is made of carbon fiber-epoxy composite and operates with 
a light gas mixture (90\% helium, 10\% isobutane). The position 
resolutions are $\sigma_{xy}\sim150\ \mu$m and $\sigma_z \sim 2$ mm. 
The momentum resolution is $\sigma(p_{\perp})/p_{\perp}\approx 0.4\%$. 
Vertexes are reconstructed with a spatial resolution of $\sim 3$~mm. 
The calorimeter~\cite{EMC} is divided into a barrel and two endcaps and 
covers 98\% of the solid angle. The modules are read out at both ends by 
photomultipliers with a readout granularity of $\sim(4.4\times 4.4)$~cm$^2$.
The arrival times of particles and the positions in three dimensions of 
the energy deposits are obtained from the signals collected at the two 
ends. Cells close in time and space are grouped into a calorimeter cluster. 
The cluster energy $E$ is the sum of the cell energies. The cluster time 
$T$ and position $\vec{R}$ are energy weighted averages. Energy and time 
resolutions are $\sigma_E/E = 5.7\%/\sqrt{E\ {\rm(GeV)}}$ and 
$\sigma_t = 57\ {\rm ps}/\sqrt{E\ {\rm(GeV)}} \oplus50\ {\rm ps}$, 
respectively.

The $U$ boson can be produced at DA$\Phi$NE through radiative decays of 
neutral mesons, such as $\phi \rightarrow \eta U$. With the statistics 
already collected at KLOE, this decay can potentially probe couplings 
down to $\epsilon \sim 10^{-3}$ \cite{Reece:2009un}, covering most of the 
parameter's range of interest for the theory. The $U$ boson can be observed 
by its decay into a lepton pair, while the $\eta$ can be tagged by one of 
its not-rare decays. 

Assuming also the existence of the $h'$, both the $U$ and the $h'$ can 
be produced at DA$\Phi$NE if their masses are smaller than M$_{\phi}$. The 
mass of the $U$ and $h'$ are both free parameters, and the possible decay 
channels can be very different depending on which particle is heavier. 
In both cases, an interesting production channel is the $h'$-strahlung, 
$e^+e^-\to Uh'$ \cite{Batell:2009yf}. Assuming the $h'$ to be lighter 
than the $U$ boson, it turns out to be very long-lived, so that the 
signature process will be a lepton pair, generated by the $U$ boson 
decay, plus missing energy. In the case $m_{h'}>m_{U}$, the dark higgs 
frequently decays to a pair of real or virtual $U$'s. In this case one can 
observe events with 6 leptons in the final state, due to the $h'$-strahlung 
process, or 4 leptons and a photon, due to the $e^+e^-\to h'\gamma$ reaction. 

Another possible channel to look for the existence of the $U$ boson is 
the $e^+e^-\to U\gamma$ process \cite{Batell:2009yf}. The expected cross 
section can be as high as $\mathcal{O}({\rm pb})$ at DA$\Phi$NE energies. 
The on-shell boson can decay into a lepton pair, giving rise to a 
$l^{+}l^{-}\gamma$ signal. 

In the following, progresses on the analyes of $\phi\to\eta U$ and 
$e^+e^-\to Uh'$ channels are reported, together with perspectives for the 
new KLOE-2 run.

\section{\boldmath The $\phi\to\eta U$ decay}

As discussed above, the search of the $U$ boson can be performed at KLOE
using the decay chain $\phi\to\eta U$, $U\to l^+ l^-$.
An irreducible background due to the Dalitz decay of the $\phi$ meson, 
$\phi\to\eta l^+l^-$, is present. This decay has been studied by SND and 
CMD-2 experiments, which measured a branching fraction of
${\mbox BR}(\phi\to\eta\,e^+e^-) = ( 1.19 \pm 0.19 \pm 0.07 )\times 10^{-4}$ 
and 
${\mbox BR}(\phi\to\eta\,e^+e^-) = ( 1.14 \pm 0.10 \pm 0.06 )\times 10^{-4}$ 
respectively \cite{phietaeeSND,phietaeeCMD2}. This corresponds to a cross 
section of $\sigma(\phi\to\eta l^+l^-)\sim 0.7$ nb, while for the signal
\begin{equation}
  \sigma(\phi\to\eta U) = \epsilon^2 \, |F_{\phi\eta}(m_U^2)|^2 \,
  \frac{\lambda^{3/2}(m_\phi^2,m_\eta^2,m_U^2)}{\lambda^{3/2}(m_\phi^2,m_\eta^2,0)}
  \, \sigma(\phi\to\eta\gamma) \sim 40\ {\rm fb} 
  \label{Eq:Reece}
\end{equation}
using $\epsilon = 10^{-3}$, $|F_{\phi\eta}(m_U^2)|^2 =1$. 
In Eq.~(\ref{Eq:Reece}), $|F_{\phi\eta}(m_U^2)|^2$ is the form factor of 
the $\phi\to\eta\gamma$ decay evaluated at the $U$ mass while the following 
term represent the ratio of the kinematic functions of the involved decays.
Despite the values of the cross section, the different shapes of the $e^+e^-$ 
invariant mass for the two processes allow to test the $\epsilon$ parameter 
down to $10^{-3}$ with the whole KLOE data set \cite{Reece:2009un}.

The best channel to search the for the $\phi\to\eta U$ process at KLOE 
is the $U\to e^+e^-$ decay for two reasons: (i) a wider range of $U$ 
boson mass can be tested; (ii) $e^\pm$ are easily identified using the 
time-of-flight (ToF) measurement. The $\eta$ can be tagged by the 
three-pion or two-photon final state, which represent $\sim85\%$ of the 
total decay rate. We have performed a preliminary analysis using the 
$\eta\to\pi^+\pi^-\pi^0$ channel, which provide a clean signal with four 
charged tracks and two photon in the final state. Studies are under way 
also for the $\eta\to\gamma\gamma$ sample.

\subsection{The $\eta\to\pi^+\pi^-\pi^0$ final state}

\begin{figure}[!t]
\includegraphics[width=0.48\textwidth]{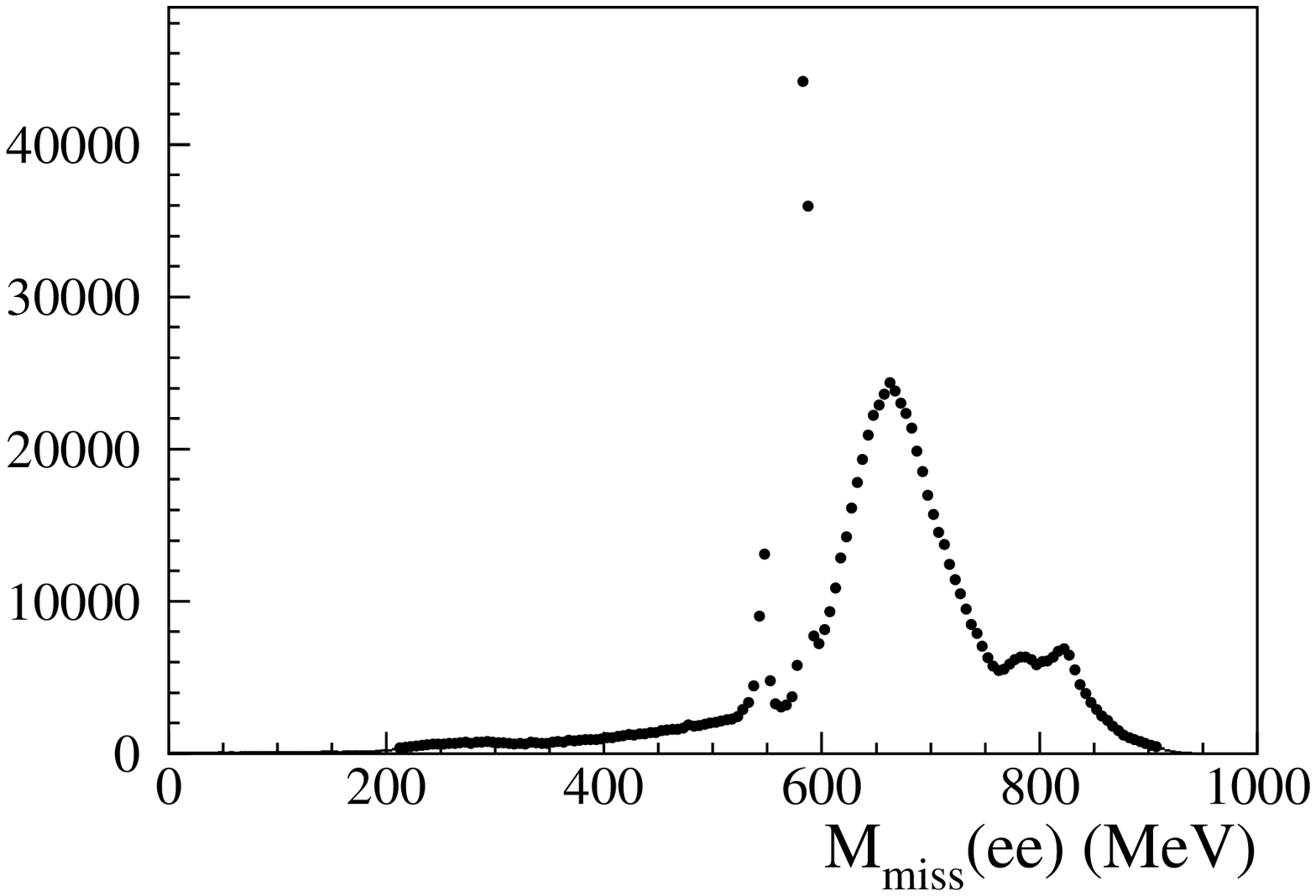}
\hfill
\includegraphics[width=0.48\textwidth]{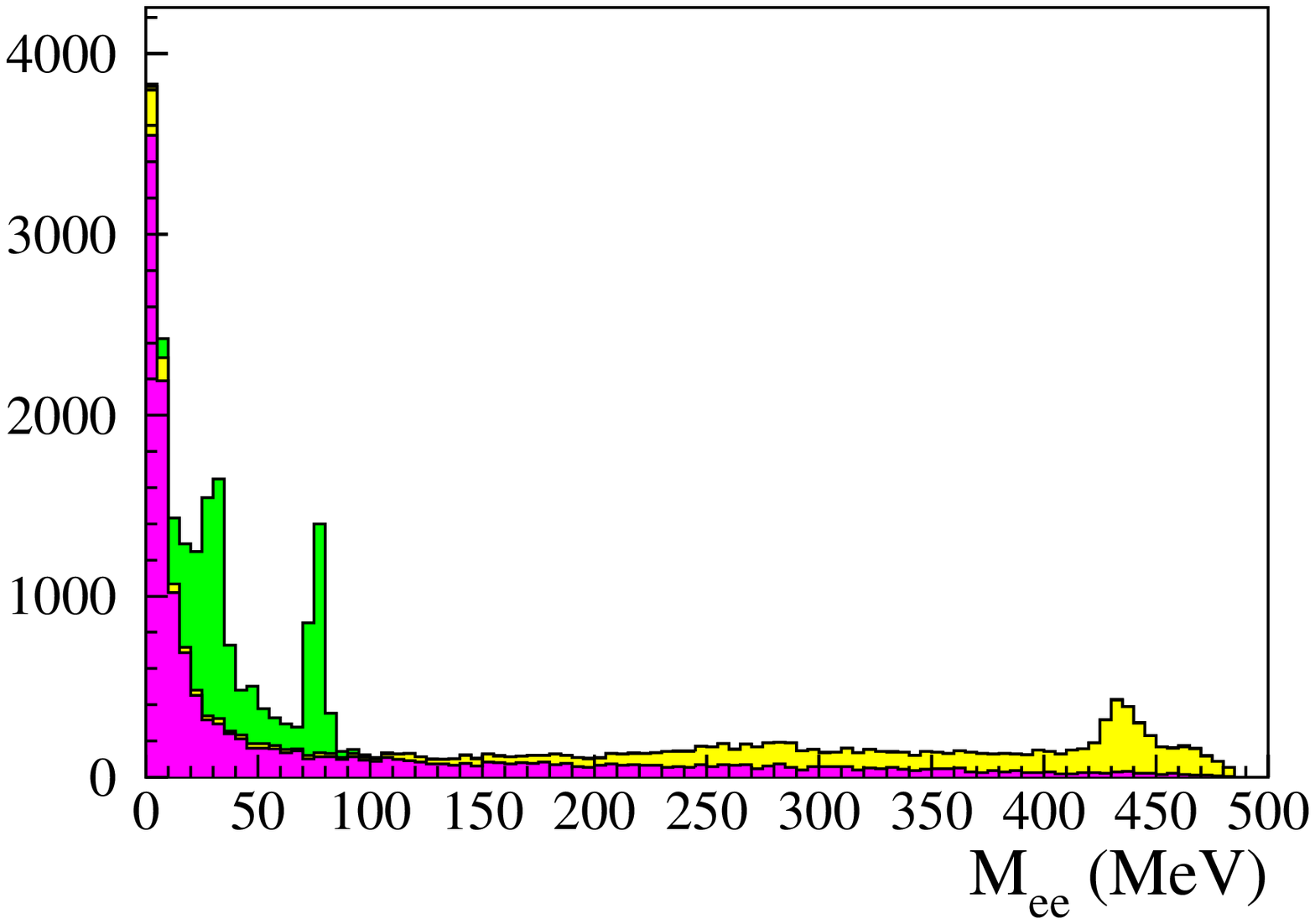}
\caption{\label{phietaee} Left: missing mass of the $e^+e^-$ pair
  for data sample after preselection cuts. The $\phi\to\eta\,e^+e^-$ 
  signal is clearly visible in the peak corresponding to $\eta$ mass.
  The second peak at $\sim 590$ MeV is due to $K_S\to\pi^+\pi^-$ 
  events with wrong mass assignment.
  Right: $M_{ee}$ distribution for data at different analysis steps: 
  preselection (green), conversion (yellow) and ToF (purple) cuts.}
\end{figure}

The preliminary analysis of the $\eta\to\pi^+\pi^-\pi^0$ final state 
has been performed on 1.5 fb$^{-1}$. 
Preselection cuts require: (i) four tracks in a cylinder around 
the interaction point (IP) plus two photon candidates; (ii) best 
$\pi^+\pi^-\gamma\gamma$ match to the $\eta$ mass using the pion hypothesis 
for tracks; (iii) other two tracks assigned to $e^+e^-$; (iv) loose 
cuts on $\eta$ and $\pi^0$ invariant masses
($495 < M_{\pi^+\pi^-\gamma\gamma} < 600$ MeV, $70 < M_{\gamma\gamma} < 200$ MeV).
These simple cuts allow to clearly see the peak due to $\phi\to\eta\,e^+e^-$
events in the distribution of the missing mass to the $e^+e^-$ pair,
$M_{\rm miss}(ee)$ (see Fig.~\ref{phietaee}-left). 
A cut $535 < M_{\rm miss}(ee) < 560$ MeV is then applied.

A residual background contamination, due to $\phi\to\eta\gamma$ events 
with photon conversion on beam pipe (BP) or drift chamber walls (DCW), 
is rejected by tracking back to BP/DCW surfaces the two $e^+$, $e^-$ 
candidates and then reconstructing the electron-positron invariant mass 
($M_{ee}$) and the distance between the two particles ($D_{ee}$). Both 
quantities are small if coming from photon conversion.
$\phi\to K \bar{K}$ and $\phi\to\pi^+\pi^-\pi^0$ events surviving analysis 
cuts have more than two pions in the final state. They are rejected using 
time-of-flight to the calorimeter. When an EMC cluster is connected 
to a track, the arrival time to the calorimeter is evaluated both with
calorimeter ($T_{\rm cluster}$) and drift chamber ($T_{\rm track}$) information.
Events with an $e^+$, $e^-$ candidate outside a 3\,$\sigma$'s window on 
the $DT = T_{\rm track}-T_{\rm cluster}$ variables are rejected.
In Fig.~\ref{phietaee}-right the $M_{ee}$ distribution evaluated at IP 
for data at different steps of the analysis is shown. The conversion and
ToF cuts remove events at low and high invariant mass values respectively.

\begin{figure}[!t]
\includegraphics[width=0.48\textwidth]{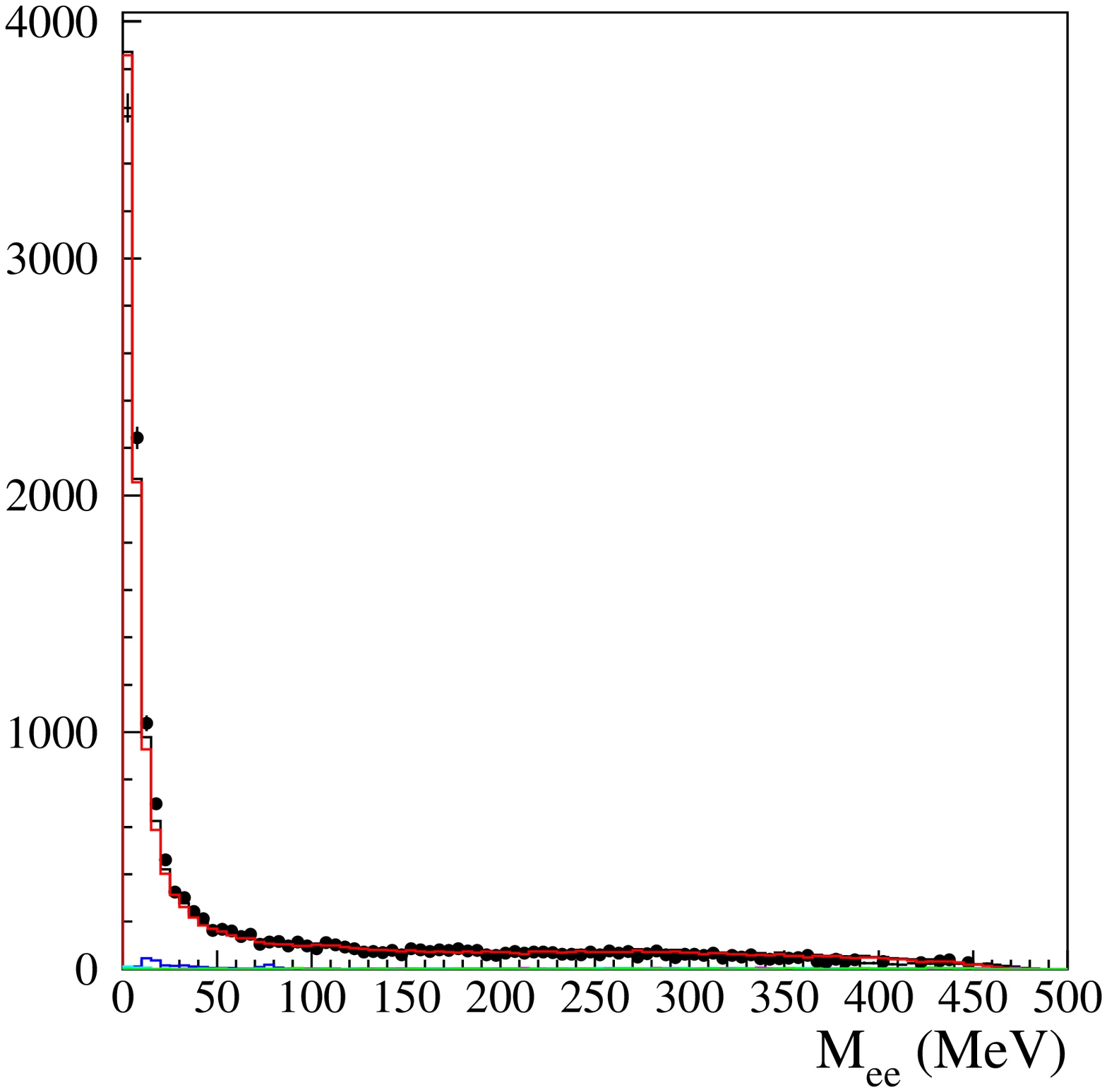}
\hfill
\includegraphics[width=0.45\textwidth]{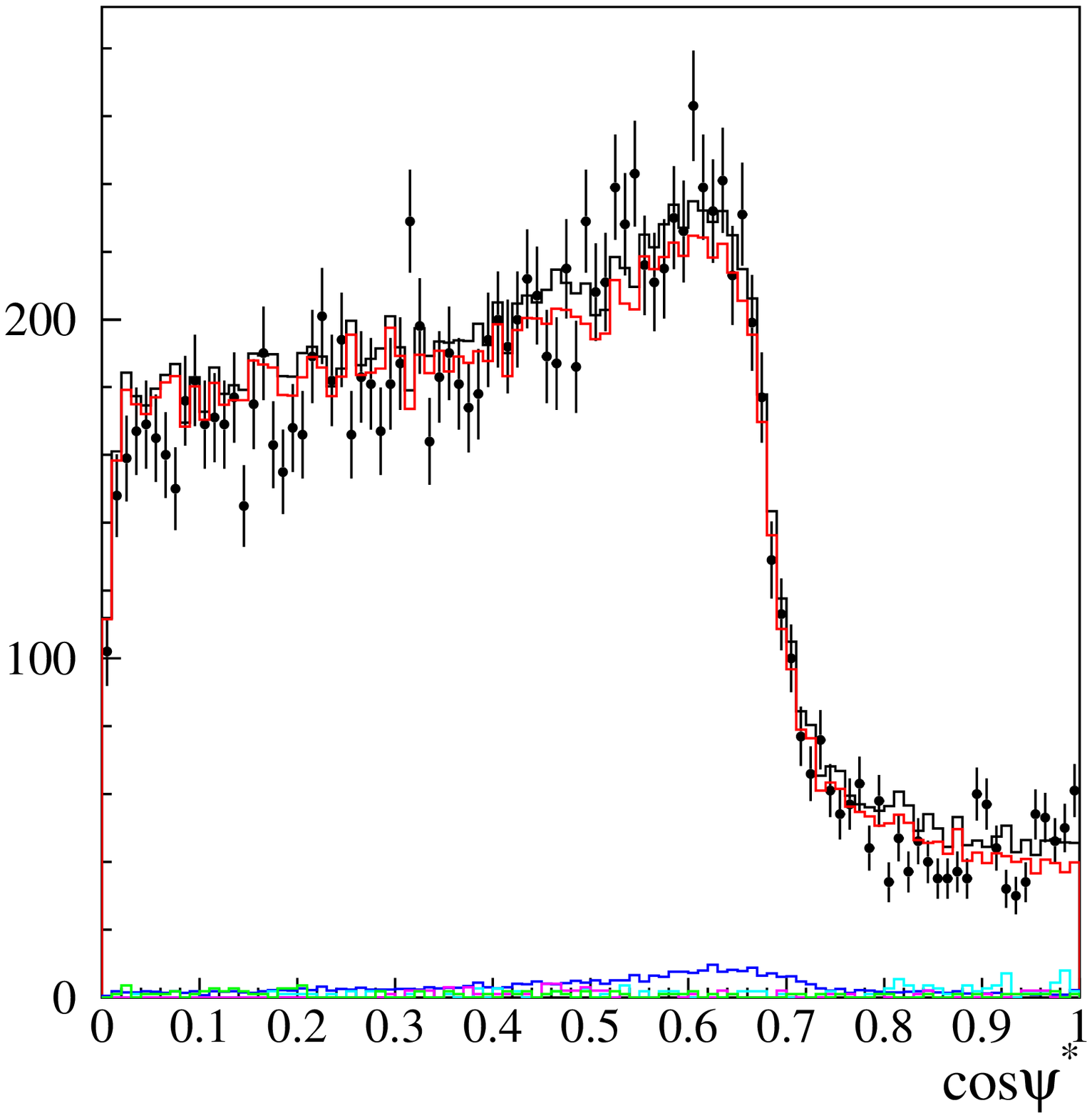}
\caption{\label{mee_cospsi} Invariant mass of the $e^+ e^-$ pair (left) 
  and $\cos\psi^*$ distribution (right) for $\phi\to\eta\,e^+ e^-$, 
  $\eta\to\pi^+\pi^-\pi^0$ events. Dots are data, the red line 
  represents the expected MC shape for signal while the residual 
  background contamination from $\phi$ decays is shown in blue. The 
  black solid line is the sum of all MC contributions.}
\end{figure}

\begin{figure}[t]
\begin{minipage}{0.45\textwidth}
\includegraphics[width=1.03\textwidth]{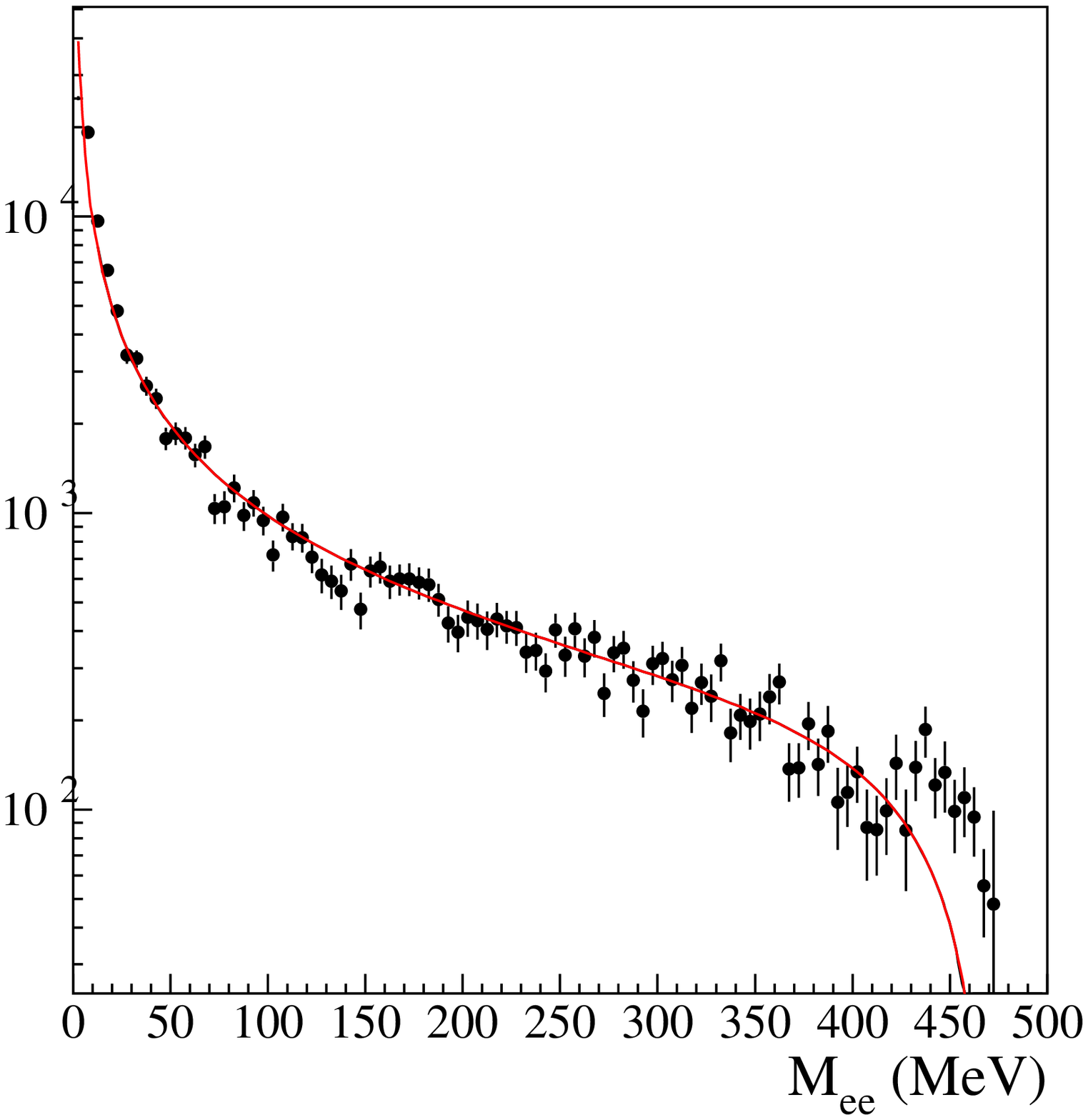}
\caption{\label{fit_mee}Fit to the $M_{ee}$ spectrum for the Dalitz decays
$\phi\to\eta\,e^+e^-$, using the $\eta\to\pi^+\pi^-\pi^0$ final state.}
\end{minipage}\hspace{0.05\textwidth}%
\begin{minipage}{0.5\textwidth}
\vspace{-0.5cm}
\includegraphics[width=1.04\textwidth]{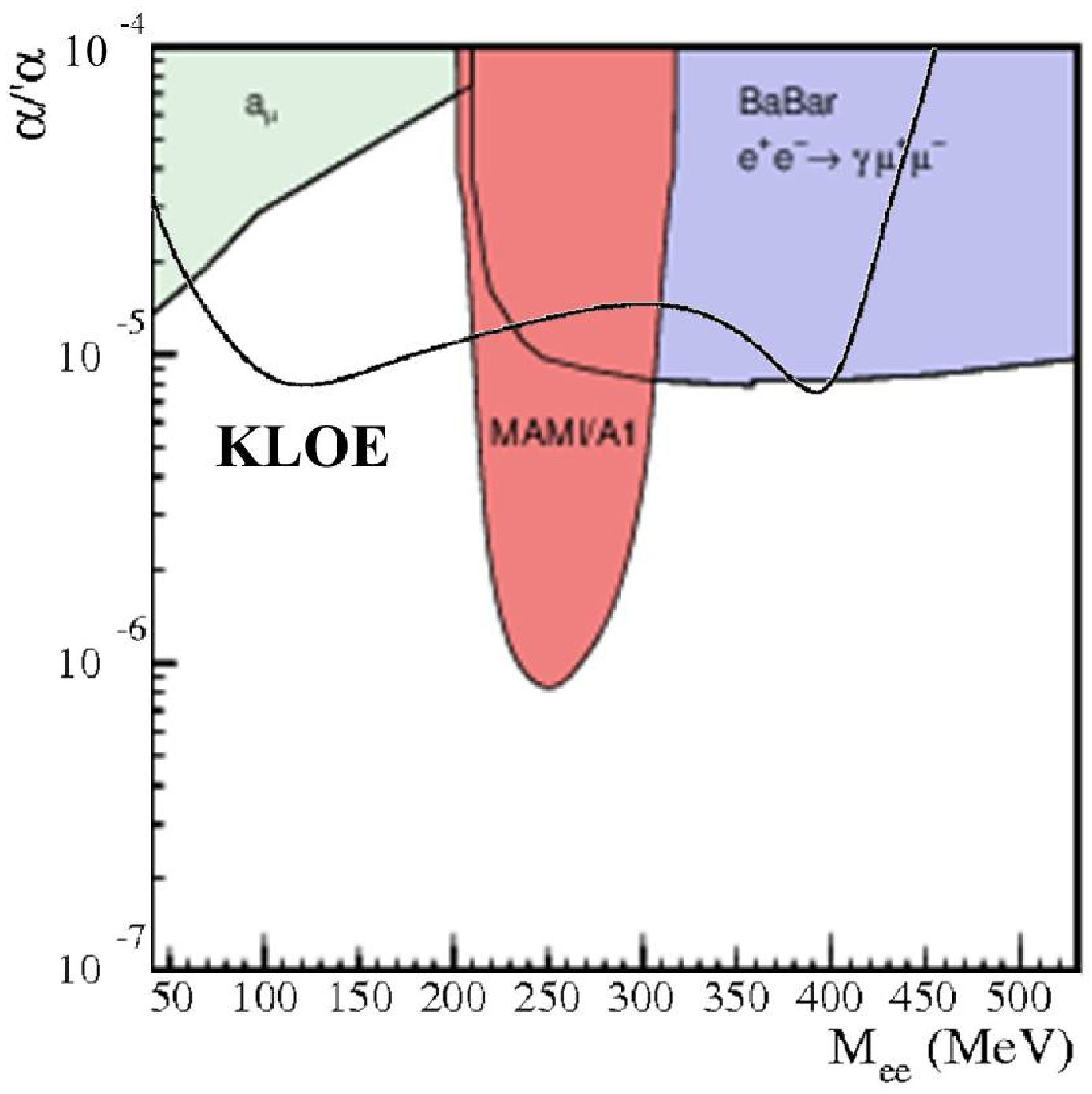}
\vspace{-0.7cm}
\caption{\label{ul}Exclusion plot for the parameter 
$\alpha'/\alpha=\epsilon^2$, compared with other existing measurements.}
\end{minipage} 
\end{figure}

In Fig.~\ref{mee_cospsi} the comparison between data and Monte Carlo (MC)
events for $M_{ee}$ and $\cos\psi^*$ distributions is shown. The second 
variable is the angle between the $\eta$ and the $e^+$ in the $e^+e^-$ 
rest frame. About 14,000 $\phi\to\eta\,e^+e^-$, $\eta\to\pi^+\pi^-\pi^0$ 
candidates are present in the analyzed data set, with a negligible residual 
background contamination. The MC description of these kind of events is 
done using Vector Meson Dominance, and the form factor slope parameter is 
taken from the measurement of SND experiment, obtained with 213 events 
\cite{phietaeeSND}. Since an accurate description of the irreducible 
background is important for the search of the $\phi\to\eta U$ signal, we 
extract it directly from our data. The decay parametrization is taken 
from Ref.~\cite{Landsberg85}:
\begin{equation}
  \frac{d}{dq^2}\frac{\Gamma(\phi\to\eta\,e^+e^-)}{\Gamma(\phi\to\eta\gamma)} = 
  \frac{\alpha}{3\pi}\frac{|F_{\phi\eta}(q^2)|^2}{q^2} 
  \sqrt{1-\frac{4m^2}{q^2}} \left(1+\frac{2m^2}{q^2}\right) 
  \left[\left(1+\frac{q^2}{m_\phi^2-m_\eta^2}\right)^2 - 
    \frac{4 m_\phi^2 q^2}{(m_\phi^2-m_\eta^2)^2} \right]^{3/2}
\end{equation}
\begin{equation}
  F(q_{\phi\eta}^2) = \frac{1}{1-q^2/\Lambda_{\phi\eta}^2}
\end{equation}
The preliminary fit to the $M_{ee}$ shape, reported in Fig.~\ref{fit_mee}, 
has $\chi^2/{\rm ndf}=2.1\%$. The resulting accuracy on form factor slope, 
$b = \Lambda_{\phi\eta}^{-2}$, is 2.8\%. Smearing matrix has not been 
included.

The $\phi\to\eta U$ MC signal has been produced according to 
Ref.~\cite{Reece:2009un}, with a flat distribution of the $e^+ e^-$ 
invariant mass. Events are then divided in sub-samples of 1 MeV width.
For each $M_{ee}$ value, signal hypothesis has been excluded at 90\% 
C.L.\ using the $\rm CL_S$ technique \cite{CLS}. For the $\phi\to\eta\,U$
signal, the  opening of the $U\to\mu^+\mu^-$ threshold has been included, 
in the hypothesis that the $U$ boson decays only to lepton pairs with
$\Gamma(U\to e^+ e^-)=\Gamma(U\to\mu^+\mu^-)$. The expected shape for the 
irreducible background $\phi\to\eta\,e^+e^-$ is obtained from our fit to 
the $M_{ee}$ distribution, taking also into account the error on number of 
background events as a function of $M_{ee}$.
In Fig.~\ref{ul} the preliminary exclusion plot on $\alpha'/\alpha=\epsilon^2$
variable is compared with recent measurements from BABAR \cite{UL_BABAR} 
and MAMI \cite{UL_MAMI} experiments. Our result greatly improves existing 
limits in the mass region $60<M_{ee}<210$ MeV, and it is also interesting 
at larger masses, where existing upper limits rely on different assumptions.

\subsection{The $\eta\to\gamma\gamma$ final state}

A similar analysis strategy has been developed for the decay chain
$\phi\to\eta U$, $U\to  e^+e^-$, $\eta\to\gamma\gamma$.
The preselection requires: (i) two tracks coming from a cylinder around 
IP, classified as $e^\pm$ using ToF information; (ii) two photon candidates; 
(iii) the cosine of the angle between photons in $\eta$ rest frame to be 
$\sim 1$; (iv) a loose cut on the total invariant mass of the system, 
$950 < M_{\rm tot} < 1150$ MeV.

The most severe background is generated by double radiative Bhabha 
scattering events and it is strongly reduced by cutting on the opening 
angle between the charged tracks and the photons. Residual non-Bhabha 
background is rejected by using further electron identification, based 
on the $E/p$ ratio for the $e^+e^-$ candidates.
The resulting background reduction is still not enough for the search of 
$\phi\to\eta U$ events.
The $M_{ee}$ spectrum obtaind with 1.7 fb$^{-1}$ shows a clear evidence 
of $\phi\to\eta\,e^+e^-$ Dalitz decays at low values and some residual 
background contamination at high $M_{ee}$ due to Bhabha events. Work 
is in progress to further improve signal to background ratio.

\section{The higgs$'$-strahlung channel}

The feasibility of the search for the process $e^+e^-\to Uh'$ has been 
done considering the $m_{h'}<m_{U}$ case. At DA$\Phi$NE energies, for 
$\epsilon\sim 10^{-3}$, a production cross section of $\approx 20$ fb is 
expected and the $h'$ has $\tau_{h'}>10\ \mu$s, escaping the detection. 
The signature is therefore a lepton pair from the $U$ boson plus missing 
energy.

The selection strategy has been optimized using Monte Carlo events. 
The signal has been generated according to Ref.~\cite{Batell:2009yf} 
in a discrete set of mass values in the range $m_U \leq 900$ MeV, 
$m_{h'}\leq 400$ MeV.
The $U\to e^+e^-$ events are not selected by any official KLOE event 
classification (ECL) algorithms, which divide the events on the basis 
of topological information and provide reconstructed data to be used 
for different analyses. On the contrary, ECL is fully efficient for 
$U\to\mu^+\mu^-$ events when $m_{h'} < 300$ MeV. 
We therefore considered the $\mu^-\mu^-$ final state only. 

Muons are identified and separated from electrons and pions using a 
neural network algorithm based on energy depositions along the shower 
depth in the calorimeter and $E/p$, $\beta$ variables. The other relevant 
cuts to reduce background contamination are: (i) missing momentum direction 
in the barrel calorimeter; (ii) a tight cut on vertex-IP distance and 
(iii) no clusters in the calorimeter, with the exception of the two 
associated to tracks. The residual background contamination is due to 
$e^+e^-\to\pi^+\pi^-\gamma/\mu^-\mu^-\gamma$ continuum events with an
undetected photon, and to $\phi\to K^+K^-\to\mu+\mu^-\nu\bar{\nu}$ with 
early decaying kaons.

\begin{figure}[!t]
\includegraphics[width=0.48\textwidth]{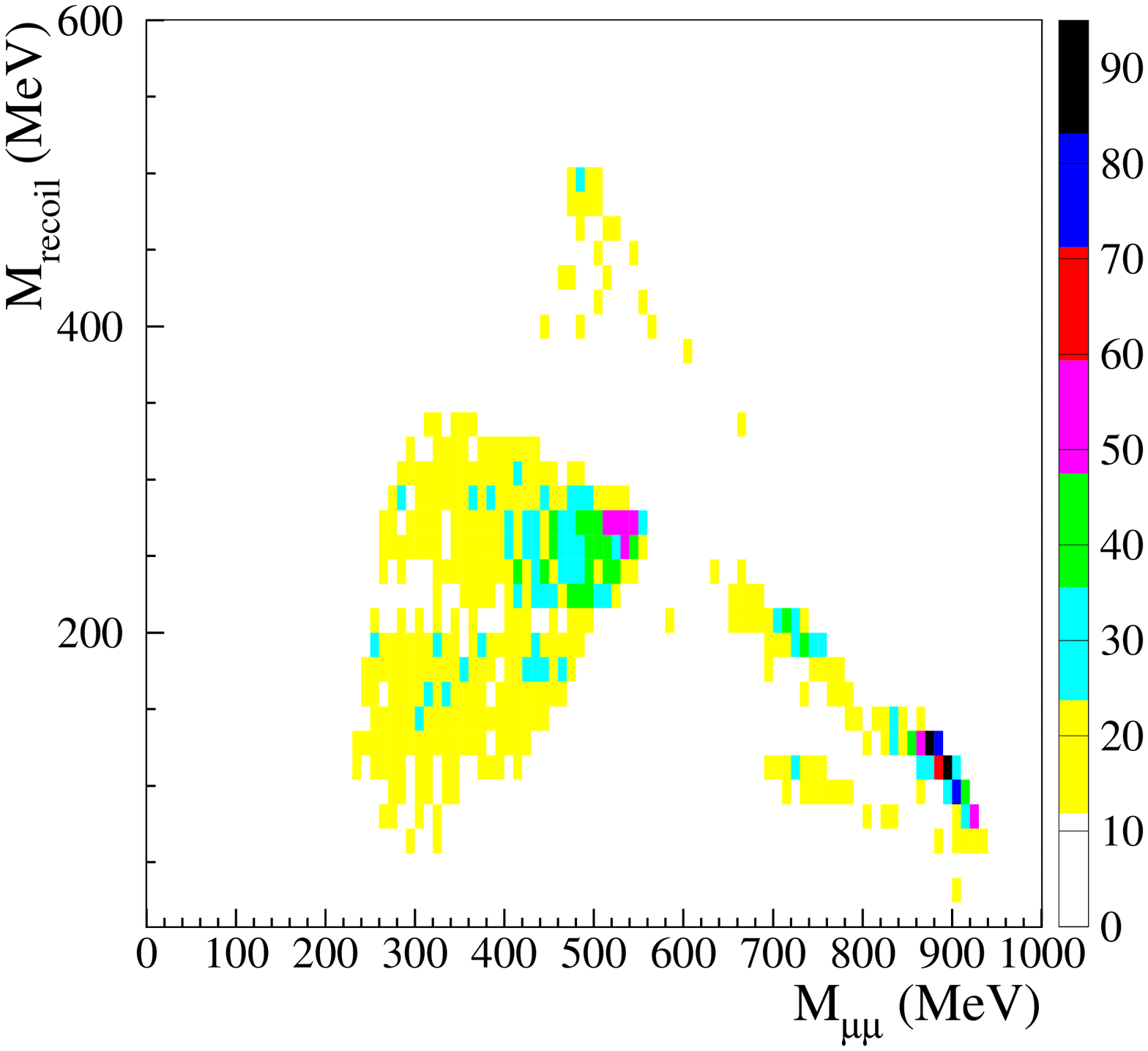}
\hfill
\includegraphics[width=0.48\textwidth]{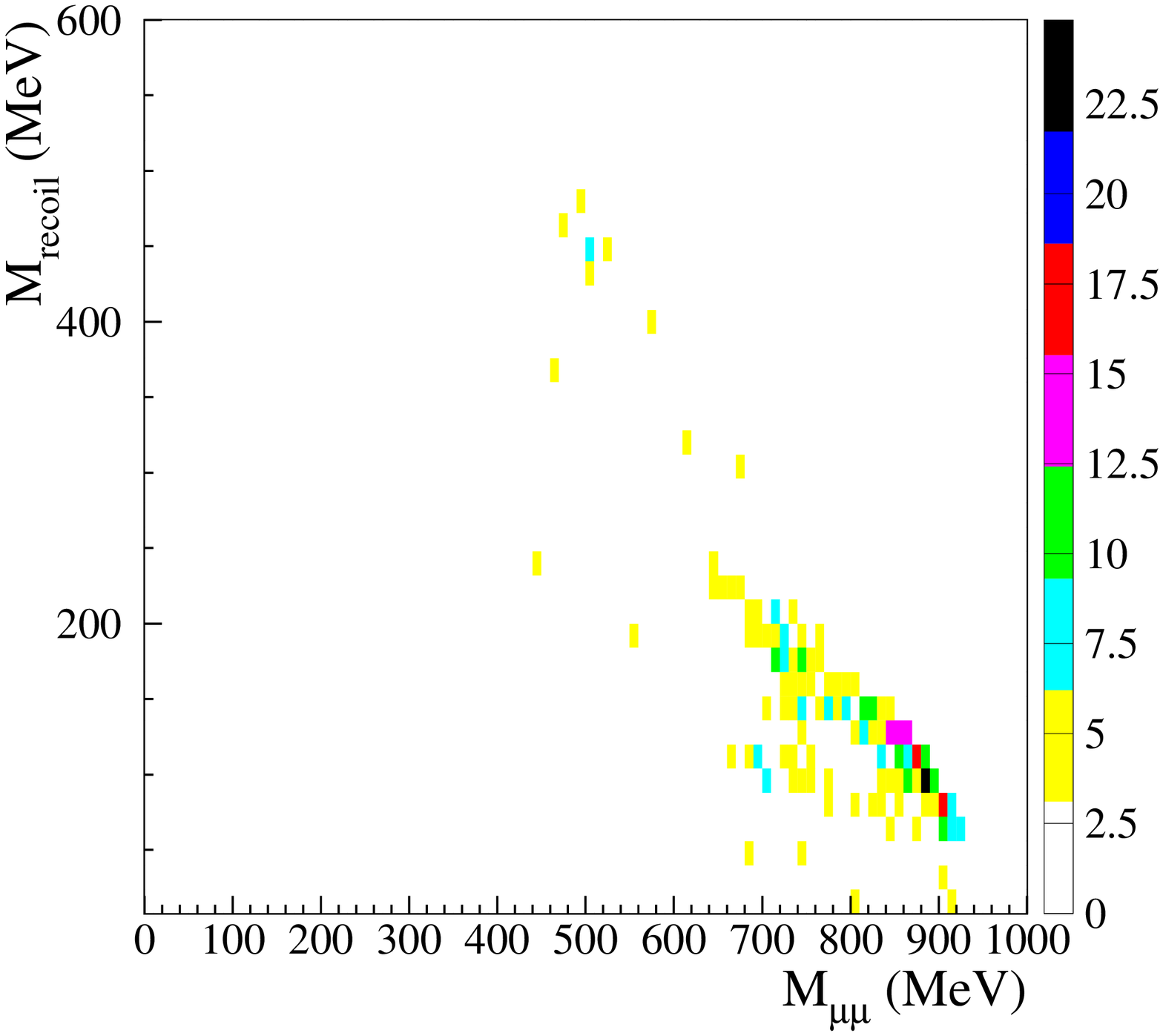}
\caption{\label{hu} Search for $e^+e^-\to h'U$, $U\to\mu^+\mu^-$, 
$h'\to ``invisible$'' events: recoil mass to the $\mu^+\mu^-$ pair as 
a function of the di-muon invariant mass for data taken at the
$\phi$ mass (left) and at $\sqrt{s}=1$ GeV (right).}
\end{figure}

In Fig.~\ref{hu}-left the distribution of the recoil mass to the 
$\mu^+\mu^-$ pair ($M_{\rm recoil}$) as a function of the di-muon invariant 
mass obtained with 1.65 fb$^{-1}$ is reported. $M_{\rm recoil}$ is evaluated 
using the center of mass energy of each run measured with Bhabha scattering
events and the momenta of the muons. Continuum background, which can be 
further reduced tuning the $\pi/\mu$ identification algorithm, is 
concentrated in the band at $M_{\mu^+\mu^-} > 700$ MeV. The $\phi\to K^+K^-$ 
channel covers a wider region of the plane ($M_{\mu^+\mu^-} < 600$ MeV, 
$M_{\rm recoil} < 300$ MeV). This background, having only two muons in the
final state and missing energy due to neutrinos, has the same signature 
of the signal.
The efficiency for $e^+e^-\to Uh'$ events is 15--40\%, depending on $m_U$, 
$m_{h'}$ masses. Taking into account the total integrated luminosity, a 
signal would show up as a sharp peak with $\leq 10$ events in the 
$M_{\rm recoil}$-$M_{\mu\mu}$ plane for $\epsilon \sim 10^{-3}$.

Being the $\phi\to K^+K^-$ background a nasty background source, we 
repeated the analysis using the off-peak sample, 0.2 fb$^{-1}$ taken
at center of mass energy of 1 GeV. As can be seen in Fig.~\ref{hu}-right,
the contribution from resonant background is not present anymore,
providing a much cleaner sample for the search of $e^+e^-\to Uh'$
candidates.

\section{Summary and perspectives for KLOE-2}

The search for $\phi\to\eta U$ with $\eta\to\pi^+\pi^-\pi^0$, using 
1.5 fb$^{-1}$ of KLOE data, results in a preliminary upper limit on
the $\alpha'/\alpha=\epsilon^2$ parameter of $\approx 1\times 10^{-5}$ 
@ 90\% C.L. in a wide $M_{ee}$ range. 
With a sample of $\mathcal{O}$(20 fb$^{-1}$) expected at KLOE-2, this 
value can be improved to $\approx 3\times 10^{-6}$ using the same $\eta$ 
decay channel. The inclusion of other final states, such as 
$\eta\to\gamma\gamma$ and $\eta\to\pi^0\pi^0\pi^0$, will further improve 
this result.

The search of the higgs$'$-strahlung channel, $e^+e^-\to Uh'$ with 
$U\to\mu^+\mu^-$ plus missing energy, is limited by a non negligible 
$\phi\to K^+K^-$ background in a wide region of the $M_{\mu^+\mu^-}$, 
$M_{\rm recoil}$ plane. Work is in progress to reduce this contribution
on the KLOE data sample. At KLOE-2, the improvement on the vertex 
resolution, achievable with the insertion of the inner tracker, will 
provide a higher rejection factor. The feasibility of a high statistics 
run at 1 GeV, where the resonant background contribution is naturally 
reduced, is also under discussion.

\section*{Acknowledgments}

We thank the DAFNE team for their efforts in maintaining low background
running conditions and their collaboration during all data-taking. 
We want to thank our technical staff: 
G.~F.~Fortugno and F.~Sborzacchi for their dedication in ensuring
efficient operation of the KLOE computing facilities;
M.~Anelli for his continuous attention to the gas system and detector
safety;
A.~Balla, M.~Gatta, G.~Corradi and G.~Papalino for electronics
maintenance;
M.~Santoni, G.~Paoluzzi and R.~Rosellini for general detector support;
C.~Piscitelli for his help during major maintenance periods.
This work was supported in part by EURODAPHNE, contract FMRX-CT98-0169; 
by the German Federal Ministry of Education and Research (BMBF) contract
06-KA-957; 
by the German Research Foundation (DFG), 'Emmy Noether Programme', 
contracts DE839/1-4;
by the EU Integrated Infrastructure Initiative HadronPhysics Project
under contract number RII3-CT-2004-506078;
by the European Commission under the 7th Framework Programme through
the 'Research Infrastructures' action of the 'Capacities' Programme,
Call: FP7-INFRASTRUCTURES-2008-1, Grant Agreement N. 227431;
by the Polish Ministery of Science and Higher Education through the
Grant No. 0469/B/H03/2009/37.

\section*{References}

\end{document}